\documentstyle[12pt,iopfts]{ioplppt}    

\newcommand{\up}{\uparrow}
\newcommand{\dw}{\downarrow}

\begin{document}

\jl{3}
\title{Energy Spectra of Few-Electron Quantum Dots}

\author{E Anisimovas\dag\ftnote{3}{Present address: Department of Theoretical 
Physics,
University of Lund, S\"{o}lvegatan 14A, S-223 62 Lund, Sweden.}
and A Matulis\ddag\ftnote{4}{Electronic mail:
matulis@uj.pfi.lt}}

\address{\dag\ Vilnius University, Sauletekio 9, 2054 Vilnius, Lithuania}

\address{\ddag\ Semiconductor Physics Institute, Gostauto 11, 2600 Vilnius,
Lithuania}

\begin{abstract}
We present the renormalized perturbation series for the energy spectrum
of the parabolic quantum dot with 2 -- 5 electrons considering ground and
the lowest excited states. The proper classification of asymptotic energy
levels is performed and behaviour of energy levels from quantum to
semiclassical regime is traced. Comparison between the present results and
those of exact numerical Hamiltonian diagonalization shows a fair accuracy
of the proposed method over the whole range of the electron-electron coupling
constant and magnetic field values. The obtained results indicate that
increasing of the number of electrons in a dot leads to more classic
behaviour of the system.

\end{abstract}

%
%
\maketitle

\section{Introduction}

Quantum dots have been studied extensively as low-dimension nanostructures
interesting from the intrinsic theoretical point of view and due to their
importance to modern electronic devices \cite{johnson95}. The most
spectroscopy experiments were carried out on quantum dots with soft parabolic
confinement potential \cite{meurer93}. Those parabolic quantum dots attract
the attention as simple model systems for electron-electron interaction
studies. Although the parabolic quantum dots have some disadvantage due to
Kohn theorem \cite{maksym90} which shows that only the center of mass motion
usually can be excited (in absorption and even in tunneling), the confining
potential in a dot very often is rather close to the parabolic one. That is
why the parabolic dots can serve as a good starting point or the reference
for the investigations of more sophisticated quantum dots. The small
nonparabolicity usually helps to reveal peculiarities of the relative electron
motion spetrum which is sensitive to the electron-electron interaction.

Only in the case of quantum dot with two electrons the exact analytical
solution of the eigenvalue problem for the parabolic dot can be obtained
\cite{taut93}. The most of theoretical investigations were mainly based
either on some mathematical tricks (which is possible in the case of
parabolic dot with 2 and 3 electrons \cite{merkt91,pfann95}), or on the
straightforward diagonalization of the Coulomb interaction \cite{haw93}.
Although the latter gives the exact solution of the problem, due to time
consuming calculations it can be applied only in the case of small number
of electrons in a dot. That is why the construction of approximate solutions
is of interest. One of them was demonstrated in \cite{maksym96} where in the
case of high magnetic field rather accurate results were obtained using the
moving Eckardt frame. Recently it was shown \cite{matulis94} that some
approximate technique based on the renormalized perturbation series can be
successfully used for the energy spectrum estimations in a wide coupling
constant range. The other technique based on Pade approximants \cite{gonzalez97}
was shown to be very successful in the case of parabolic dot with small number
of electrons. Both of those approximations are based on some interpolating
technique between the perturbation series in coupling constant and the
asymptotic expansion valid in large coupling constant range.

In the present paper in order to obtain the total picture of energy spectrum,
including the ground and some excited states, we performed the proper 
classification
of the asymptotic energy levels paying attention to the symmetry properties of
the spin wave function part and constructed the renormalized perturbation series
for the energy of quantum dots with 2 -- 5 electrons based on rahter simple two
term expansion in both limiting regions of the coupling constant.

The present paper is organized as follows. In section 2 the problem is formulated
and the perturbation series technique is outlined. The asymptotic series are
formulated in section 3 and the spin wave function part for asymptotic region
is constructed in section 4. The interpolating technique is described in section 5
and in the last section 6 the results for quantum dots with 2 -- 5 electrons
are presented and discussed. In appendix A the algebraic expression for Coulomb
matrix element is given and in appendix B all coefficients which are necesssary 
for
construction of the renormalized perturbation series are collected.

\section{Perturbation series}

Generally the system of $N$ electrons in a quasi-2D quantum dot
is described by the following Hamiltonian
\begin{equation}\label{basic_ham}
  H = \sum_{i=1}^{N}\left[ -\frac{\hbar^2}{2m}\nabla_i^2 +
  \frac{m\omega_0^2}{2} {\bi r}_i^2 \right] +
  \sum_{i<j}^{N} \frac{e^2}{\epsilon |{\bi r}_i-{\bi r}_j|}.
\end{equation}
Here $m$ stands for the effective electron mass, $\epsilon$ is the static
dielectric constant,  and $\omega_0$ is the characteristic frequency of
parabolic confinement potential. For the sake of convenience we shall scale
the coordinates ${\bi r} \to a_0{\bi r}$ ($a_0=\sqrt{\hbar/m\omega_0}$) and
measure the energies in dimensionless units of $\hbar \omega_0$. After this
transformation we obtain the dimensionless Hamiltonian
\begin{equation}
\label{bedim}
  H=\frac{1}{2}\sum_{i=1}^{N}\left( -\nabla_i^2 + {\bi r}_i^2 \right)
  + \lambda \sum_{i<j}^{N} \frac{1}{|{\bi r}_i-{\bi r}_j|}
  = H_0+\lambda V
\end{equation}
with the single dimensionless coupling constant $\lambda=a_0/a_B$
($a_B=\epsilon\hbar^2/me^2$ is the effective Bohr radius) characterizing
the energy spectrum of electrons in the parabolic quantum dot.

The simplest way to solve the eigenvalue problem is just to use the perturbation
series in coupling constant $\lambda$
\begin{equation}
\label{perturb}
  E(\lambda)=E_0+E_1 \lambda.
\end{equation}
The first coefficient is the eigenvalue of the unperturbed Hamiltonian $H_0$ and
can be presented as the sum of noninteracting electron energies
\begin{equation}\label{one}
  E_0=\sum_{i=1}^N \varepsilon(i), \qquad \varepsilon(i)=1+|m_i|+2n_i
\end{equation}
where $i=(n_i,m_i,s_i)$ labels the one electron quantum number set.

The first order correction $E_1$ can be estimated in a standard way (see, for
instance \cite{fjaer97}) and presented as a sum of two-electron Coulomb matrix
elements
\begin{equation}\label{meleme}
\fl  \langle ij|V|j'i'\rangle = \delta_{s_i,s_{i'}}\delta_{s_j,s_{j'}}
  \int\!d^2r_1\!\int\!d^2r_2 \,\frac{1}{|{\bi r}_i-{\bi r}_j|}
  \phi^*(i|{\bi r}_1)\phi^*(j|{\bi r}_2)\phi(j'|{\bi r}_2)\phi(i'|{\bi r}_1)
\end{equation}
where the one electron wave functions can be expressed in associated Laguerre
polynomials
\begin{equation}\label{wf}
  \phi(n,m|r,\varphi)=\sqrt{\frac{2n!}{2\pi(|m|+n)!}}\,e^{im\varphi}
  r^{|m|}e^{-r^2/2}L_n^{|m|}(r^2).
\end{equation}
For numerical estimation of the Coulomb matrix element (\ref{meleme}) we used
the analogue of Girvin and Jach expression \cite{girvin83} which is given
in \ref{aps1}.

When calculating the first order correction the main problem is the degeneracy
of zero order energy levels because the good quantum numbers of the total angular
momentum $M$ and the total spin $S$ are not sufficient for classification of
energy levels. That is why we used numerical diagonalization of the degenerate
blocks of the above first order correction matrix. The obtained results -- the
coefficients $E_0$ and $E_1$ are collected in \ref{coeff}.

\section{Asymptotic expansion}

In the $\lambda \to \infty$ case the kinetic energy of the electron system
is small if compared to the potential one. It is known that such systems are
strongly correlated and show tendency for Wigner crystallization \cite{maks93}.
Monte Carlo results for classical particles in the parabolic dot
\cite{bedanov94} show that the particles crystallize in some structure of
concentric rings. When the number of electrons in a dot is small $N \le 5$
they fall into a single ring on which the electrons are located equidistantly
and separated by the angle $\alpha=2\pi/N$. The quantum dynamics of such a Wigner
crystal can be described by means of the rotation and vibration modes. The general
theory of vibration modes is given in \cite{maksym96}. It is based on introducing
the moving Eckardt frame and the diagonalization of the vibrational Hamiltonian
in the harmonic approach. In the case of small number of electrons located on
a single ring the vibrational Hamiltonian and the problem of its diagonalization
can be essentially simplified by representing the electron coordinates as complex
variables $z=x+iy$ and turning them to the local electron coordinates on the ring
by means the following transformation
\begin{equation}\label{coord}
  z_n \to (R+z_n)\e^{i\alpha n}.
\end{equation}
Here $n$ labels the electrons on the ring. We shall follow the simpler way
restricting our consideration by the following two term asymptotic expansion
\begin{equation}
\label{eksp}
  E(\lambda)=c_0\lambda^{2/3}+c_1.
\end{equation}
The first term of it can be easily obtained minimizing the potential in 
(\ref{bedim}).
The minimization leads to the equilibrium ring radius $R=(\lambda A)^{1/3}$
(where $A=\sum_{n=1}^{N-1}|4\sin(\alpha n/2)|^{-1}$) which corresponds to the main
term $c_0 \lambda^{2/3}=1.5NR^2$ in the asymptotic energy expansion.
The next term in expansion (\ref{eksp}) corresponds to the ring oscillations.
To obtain it we insert (\ref{coord}) into Hamiltonian (\ref{bedim}) and expand the
potential into $z$-series. It leads to the following second order correction to
the equilibrium potential
\[
  V_2=\frac{1}{2}\sum_{n=1}^N |z_n|^2+\frac{N}{32A}\sum_{n=1}^{N-1}
  (3\gamma_n^2+3\gamma_n^{*2}+2|\gamma_n|^2)
\]
where
\[
  \fl  \gamma_{nm}=\frac{z_n \e^{i\alpha n}-z_m \e^{i\alpha m}}
  {\e^{i\alpha n}-\e^{i\alpha m}}
  =\frac{z_n\left[1-\e^{i\alpha(n-m)} \right]+z_m\left[1-\e^{i\alpha(m-n)}
  \right]}
  {4\sin^2[\alpha(n-m)/2]} \equiv \gamma_{n-m}.
\]
Now let us apply the Fourier transform
\begin{equation}\label{fourier}
  z_n=\frac{1}{\sqrt{N}}\sum_k \e^{ik\alpha n}z_k
\end{equation}
where $k=0,1,\dots,N-1$ denotes the vibration modes. Inserting it into the above
potential expression and subsequently into (\ref{bedim}) after simple but 
laborious
transformation we obtain the following expression for the asymptotic Hamiltonian
with the second order potential corrections
\begin{equation}\label{hamz}
  H=\sum_k H_k = \sum_k \left\{ -2\frac{\partial^2}{\partial z_k \partial
  z_k^{*}}+
  A_k z_kz_k^{*}+B_k(z_kz_{-k}+z_k^{*}z_{-k}^{*}) \right\}
\end{equation}
where
\begin{eqnarray}
  A_k =\frac{1}{2}+\frac{1}{16A}\sum_{n=1}^{N-1}|\sin(\alpha n/2)|^{-3}
       \sin^2[(k+1)\alpha n/2], \nonumber \\
  B_k =\frac{3}{32A}\sum_{n=1}^{N-1}|\sin(\alpha n/2)|^{-3}
       [\sin^2(\alpha n/2)-\sin^2(k\alpha n/2)]. \nonumber
\end{eqnarray}

The Hamiltonian is composed of parts $H_k$ which describe the behaviour of both
polarization of $\pm k$ modes as a system of four coupled harmonic oscillators.
They can be diagonalized in a standard way just introducing the proper variables.
Let us separate the real and imaginary parts of the initial variables, and 
introduce
the new variables in the following way
\begin{equation}
  z_k=x+iy, \qquad z_{-k}=u+iv.
\end{equation}
In these variables every Hamiltonian (\ref{hamz}) part $H_k$ splits into two
terms. The first term is
\begin{equation}\label{pirmas}
  H_k^{(1)}=-\frac{1}{2}\left(\frac{\partial^2}{\partial x^2}+
  \frac{\partial^2}{\partial u^2}\right)+A_k x^2+A_{-k} u^2+
  2(B_k + B_{-k}) xu
\end{equation}
while the other part $H_k^{(2)}$ can be obtained just replacing $x\to y$ and $u\to
-v$.

Both $H_k^{(1,2)}$ have the same eigenvalues that can be found by the
diagonalization of the symmetric matrix
\[
  \left( \begin{array}{cc} A_k & B_k+B_{-k} \\
  B_k+B_{-k} & A_{-k} \end{array} \right)
\]
which defines the quadratic form of the potential term, and leads to the following
expression for the eigenfrequencies
\begin{equation}\label{klasik}
  \omega_k^{(\pm)} =\sqrt{A_k+A_{-k} \pm \sqrt{[A_k-A_{-k}]^2+4[B_k+B_{-k}]^2}}.
\end{equation}

Thus, since the transformed Hamiltonian corresponds to the set of independent
harmonic oscillators the corresponding vibration wave function $\Phi$ can be
constructed in a standard way introducing the ``phonon'' creation 
$a_{k,\beta}^{+}$
and annihilation $a_{k,\beta}$ operators
\begin{equation}\label{wfvib}
  \Phi=a_{k_1,\beta_1}^{\dagger}\cdots a_{k_p,\beta_p}^{\dagger}|0\rangle.
\end{equation}
The ``vacuum'' state $|0\rangle$ corresponds to the system ground state,
and the symbol $\beta=1,2$ stands for polarization.

For the interpolation purposes it is important to know the wave function symmetry
properties with respect to the rotation of the electronic ring as a whole. Let us
denote this operation by operator ${\cal P}$. It is evident that the ground state
is invariant under that operation, i.~e. ${\cal P}|0\rangle=|0\rangle$. As the
above rotation is equivalent to the substitution $n\to n+1$ it follows from 
expression
(\ref{fourier}) that ${\cal P}z_k=\exp(ik)z_k$. Due to the mixing of both $\pm k$
modes the operators $a_{k,\beta}^{\dagger}$ and wave function (\ref{wfvib}) obey
more complicated symmetry condition. However, because of the degeneracy of 
obtained
vibration states it can be essentially simplified using the proper superposition 
of
the eigenstates of both Hamiltonian parts $H_k^{(1,2)}$. So, we shall assume that
the wave function superposition of this sort is constructed, and consequently the
phonon operators obey the symmetry condition
\begin{equation}\label{opsym}
  {\cal P}a_{k,\alpha}^{\dagger}=\e^{-ik}a_{k,\alpha}^{\dagger}.
\end{equation}

We shall use the above symmetry condition for the classification of obtained
asymptotic vibration modes whose frequencies $\omega_k^{(\beta)}$ are collected
in \tref{asfreq}.
%
%
\begin{table}
\caption{The ring vibration frequencies and the asymptotic expansion
         coefficients}
\footnotesize\rm
\begin{tabular}{@{}llllllllll}
\br
  & N & \multicolumn{2}{l}{2} & \multicolumn{2}{l}{3} & \multicolumn{2}{l}{4}
  & \multicolumn{2}{l}{5} \\
\ns
  &\crule{9}\\
k  & $\beta$ & $1$ & $2$ & $1$ & $2$ & $1$ & $2$ & $1$ & $2$ \\
\mr
0 && 1.7321 & 0.0\phantom{321} & 1.7321 & 0.0 & 1.7321 & 0.0 & 1.7321 &
0.0\phantom{321} \\
1 && 1.0    & 1.0 & 1.2247 & 1.0    & 1.3186 & 1.0    & 1.3719 & 1.0    \\
2 &&        &     & 1.2247 & 1.0    & 1.4888 & 0.8852 & 1.6890 & 0.7274 \\
3 &&        &     &        &        & 1.3186 & 1.0    & 1.6890 & 0.7274 \\
4 &&        &     &        &        &        &        & 1.3719 & 1.0    \\
\mr
& $c_0$       & 1.1905     && 3.1201  && 5.8272 && 9.2801 &   \\
& $c_1^{(0)}$ & 1.8660     && 3.0908  && 4.3716 && 5.6544 &   \\
\br
\end{tabular}
\label{asfreq}
\end{table}
%
%
Two modes with eigenfrequencies $\omega_k^{(\beta)}=1$ correspond to the center
of mass motion. We do not take them into account as we consider the electron
relative motion energy spectrum only. The mode $k=0$ corresponds to the 
`breathing'
mode for $\beta=1$ and to the rotation of the ring as a whole for $\beta=2$. The
last case should be considered separately. Actually, in the harmonic approximation
used the rotation mode is separated from the vibration modes. Its eigenfunction is
$\exp(iM\varphi)$ ($\varphi$ is an angle determining rotation of the ring as a 
whole)
and the corresponding energy is $M^2/2NR^2 \sim \lambda^{-2/3}$. It is seen that 
in
the case of large angular momenta $M$ this term becomes essential and thus it 
should
be included into the main asymptotic energy term, and consequently it leads to the
equilibrium ring radius dependence on the total angular momentum $M$ (see
\cite{maksym96} for instance). However, the above term is of order 
$\lambda^{-2/3}$
and lies beyond the (\ref{eksp}) approximation. Hence considering the states with
small $M$ only we neglect this rotation mode correction (and the ring radius
dependence on $M$ as well) together with the anharmonicity corrections which are 
of
the same order of magnitude. We should keep in mind, however, that there are many
states labeled by different $M$ values corresponding to every asymptotic vibration 
term.

Thus, the first order correction is given by
\[
  c_1 = c_1^{(0)} + \sum_{k,\beta} n_{k,\beta} \omega_k^{(\beta)}
\]
where $c_1^{(0)}$ is the contribution of the ground vibration state energy and
the symbol $n_{k,\beta}$ stands for the filling factor of the corresponding
phonon mode. Those coefficients together with first term coefficients $c_0$
are given in the last two lines of \tref{asfreq}.

Separating the rotation mode we shall represent the total wave function in the
asymptotic region as a product of rotation, vibration and spin parts
\begin{equation}\label{prod}
  \Psi = \e^{iM\varphi}\Phi \Upsilon.
\end{equation}
According to the Pauli's principle the total wave function should be 
antisymmetric.
In the asymptotic region the classical electron equilibrium positions are 
separated \
by high potential barriers. That is why we shall restrict ourselves only to the
physically important permutation --- the rotation ${\cal P}$. As the rotation
corresponds to the odd permutation in the case of the even number of electrons
(and vice versa) the total wave function obeys the following symmetry condition
\begin{equation}\label{psisym}
  {\cal P}\Psi = (-1)^{N-1}\Psi .
\end{equation}
According to (\ref{wfvib}--\ref{prod}) the rotation and vibration parts
obey the condition
\begin{equation}\label{vibsym}
  {\cal P}\e^{iM\varphi}\Phi = \e^{i\alpha(M-\gamma)}\e^{iM\varphi}
     \Phi, \qquad \gamma = {\sum_{k,\beta}}'
   kn_{k,\beta}.
\end{equation}
The prime in summation indicates that the mode corresponding to the ring
rotation is excluded. The different factors in the two above conditions should
be compensated by the proper choice of the spin function part.

\section{Spin wave function part}

Following \cite{maksym96} we diagonalize
the ${\cal P}$ operator in the $S_z$ (total spin $z$-projection operator)
eigenfunction space and construct the spin function part obeying the
symmetry condition
\begin{equation}\label{spinsym}
  {\cal P}\Upsilon = \e^{i\alpha\tau}\Upsilon
\end{equation}
which together with (\ref{psisym}), (\ref{vibsym}) leads to the following
selection rule
\begin{equation}
  M+\tau-\gamma = Np.
\end{equation}
In the case of odd number $N$ of electrons in a dot the symbol $p$ stands for
an arbitrary integer while in the case of the even $N$ it
should be replaced by $p + 1/2$. The integer parameter $\tau$ which characterizes
the spin function part symmetry properties should be defined for every particular
case. In the case when the above procedure leads to the degeneracy of spin states
the total spin operator $S^2$ should be diagonalized additionally.

The case of $2$ electrons is a trivial one. There is the triplet function
$|\!\up\up\rangle$ with $S_z=1$ corresponding to $\tau=0$ and the singlet function
$\{|\!\up\dw\rangle - |\!\dw\up\rangle \}/\sqrt{2}$ with $S_z=0$ corresponding
to $\tau=1$. We shall not consider the functions with $S_z < 0$ as they give
no extra information.

The case of 3 electrons is considered in detail in \cite{maksym96}. There
is quartet state corresponding to $\tau=0$ and two doublet states with
$\tau=\pm 1$.

Applying the same treatment to the system with 4 and 5 electrons we obtain the
spin function parts with the symmetry properties of (\ref{spinsym}) type. All
the necessary $\tau$ values are collected in \tref{spin}.
%
\begin{table}
\caption{$\tau$ values for the various multiplets and the ground state terms.}
\footnotesize\rm
\begin{tabular}{@{}lccccccc}
\br
N  & $S=0$ & $S=1/2$ & $S=1$ & $S=3/2$ & $S=2$ & $S=5/2$ & Terms \\
\mr
2 & $1$   & -- & $0$ & -- & -- & -- & $^1S$, $^3P$, $^1D$, $^3F$, $\cdots$ \\
3 & --     & $1,2$ & -- & $0$ & -- & -- & $^4S$, $^2P$, $^2D$, $^4F$, $\cdots$ \\
4 & $0,2$ & --  & $1,2,3$  & -- & $0$ & -- & $^{1,3}S$, $^3P$, $^{1,5}D$, $^3F$,
$^{1,3}G$, $\cdots$ \\
5 & -- & $0,1,2,3,4$ & -- & $1,2,3,4$ & -- & $0$ & $^{2,6}S$, $^{2,4}P$,
$^{2,4}D$, $^{2,4}F$, $^{2,4}G$ $\cdots$ \\
\br
\end{tabular}
\label{spin}
\end{table}
%
%

The above selection rule gives us the possible terms for every vibration state
characterized by $\gamma$ value. The terms for the ground vibration state
($\gamma=0$) are collected in the last column of \tref{spin}. For the sake of
brevity we use the notation analogous to that adopted in atomic spectroscopy.
Thus the capital letters $S, P, D, \cdots$ indicate the corresponding
$M=0,1,2,\cdots$ values, and the left superscript symbol stands for multiplicity
($2S+1$). The terms corresponding to the excited vibration states with $\gamma \ne 
0$
can be easily obtained from those presented in \tref{spin} applying the shift
$M\to M+\gamma$. Further we shall label those excited terms with the additional
right underscript symbol $k$ indicating the modes of the phonons present in
that state.

\section{Interpolation}

In constructing the interpolating expression we shall follow \cite{matulis94}
where the perturbation series renormalization procedure is given in details.
The main idea is replacing of (\ref{bedim}) by the generalized Hamiltonian
\begin{equation}\label{duparam}
  H=\frac{1}{2}\sum_{i=1}^{N}\left( -\xi^2\nabla_i^2 + {\bi r}_i^2 \right)
  + \lambda \sum_{i<j} \frac{1}{|{\bi r}_i-{\bi r}_j|},
\end{equation}
and using the scaling relation for the generalized eigenvalue
\begin{equation}
  E(\xi,\lambda) = \xi E(1,\lambda\xi^{-3/2}).
\end{equation}
The eigenvalue of the basic problem with Hamiltonian (\ref{bedim}) is given
as $E(\lambda)=E(1,\lambda)$.

The scaling relation defines the trajectory $\lambda=\lambda_0 \xi^{3/2}$
in the $\xi\lambda$-plane as it is shown in \fref{fig1}. That trajectory
enables to map the vertical dashed line $\xi=1$ (where we need to obtain the
eigenvalue of the basic problem) onto tilted line $KP$ which we define as
\begin{equation}\label{mapline}
\begin{array}{ll}
  \lambda = K\beta, & K=(1+\tan\phi)/2 \\
  \xi = P(1-\beta), & P=(1+\cot\phi)/2 .\\
\end{array}
\end{equation}
For instance, the point $\lambda_0$ is mapped to the point $(\xi,\lambda)$.

Contrary to line $\xi=1$ the line $KP$ is located in the finite region of the
$\xi\lambda$-plane ($0<\beta<1$), and that is why the energy $E(\xi(\beta),
\lambda(\beta))$ can be successfully expanded into $\beta$-series. Adjusting
the above series to the $\lambda$-series (\ref{perturb}) in the region
$\lambda\to 0$ and to the asymptotic expansion (\ref{eksp}) in the region
$\lambda \to \infty$ we obtain the final
expression for the renormalized series
\begin{eqnarray}\label{renorm}
  E &=& (b_0+b_1\beta+b_2\beta^2+b_3\beta^3)/(1-\beta) \nonumber \\
  \lambda &=& \frac{K\beta}{P^{3/2}(1-\beta)^{3/2}}.
\end{eqnarray}
The expansion coefficients can be expressed in terms of those in (\ref{perturb})
and (\ref{eksp}) as
\begin{eqnarray}\label{rncoef}
  b_0 &=& E_0 \nonumber \\
  b_1 &=& KP^{-3/2}E_1-E_0 \nonumber \\
  b_2 &=& 7/3\,K^{2/3}P^{-1}c_0+c_1-E_0-2KP^{-3/2}E_1 \nonumber \\
  b_3 &=& -4/3\,K^{2/3}P^{-1}c_0-c_1+E_0+KP^{-3/2}E_1.
\end{eqnarray}

As we restricted ourselves with the two term approximation in both (\ref{perturb})
and (\ref{eksp}) expansions we introduced the additional adjustable parameter
$\phi$ (as proposed in \cite{fernandez84}) for improving the interpolating
accuracy. The geometric meaning of that parameter is clear from \fref{fig1} ---
it indicates the rotation angle of the mapping line $KP$ around the point $M$
to which the energy $E(\sqrt{2})$ is always mapped. The rotation changes
the relative contributions of the $\lambda$-series and the asymptotic
expansion to the renormalized series. The value $\phi=0$ corresponds to the full
control of the renormalization procedure by the $\lambda$-series and the value
$\phi=\pi /2$ to that of the asymptotic expansion.

It was shown in \cite{fernandez84} that usually the energy exhibits some plateau
as a function of the parameter controlling the mapping line ($\phi$ in our case).
When the number of terms in the $\lambda$-series is increased that plateau shows
the tendency to increase, too, thus giving a good reason for the controlling
parameter choice. In \fref{fig2} the energy at the fixed point $\lambda=\sqrt{2}$
as a function of the parameter $\phi$ is shown by the solid line in the case of
the first excited triplet state of the two electron system. As our 
$\lambda$-series
are rather short we have no prominent plateau. However, some curve twist is
present, and we hope that it is a rather good idea to choose the controlling
parameter $\phi$ somewhere between the Max and Min points. The difference of
energy values corresponding to those points can serve as an estimation of the
interpolation accuracy. We see that the accuracy is of about 1\%.

We used the inflection point of $E(\sqrt{2})$ versus $\phi$ curve for the
controlling parameter $\phi$ choice. The energy value obtained with that choice
is shown by dashed line in \fref{fig2}. Comparing it with the exact result which
is known for two electron case (dotted line) we see that the accuracy is even 
better
than 1\%. The same energy versus $\phi$ behaviour and the same accuracy was 
obtained
for all other considered terms. The inflection
point values of $\phi$ for all the considered terms are given in \ref{coeff} in 
tables
\ref{tap1} and \ref{tap2}. They together with coefficients $E_0$, $E_1$ and
expressions (\ref{mapline}--\ref{rncoef})
define the renormalized perturbation series and enable to obtain the energies in
the whole range of $\lambda$ values.

We should note that according to \cite{matulis94} the same parameter values
can be used for constructing the renormalized series in the case when the
additional magnetic field $B$ is applied in the direction perpendicular to
the quantum dot plane. In that case we have
\begin{equation}\label{energy_mag}
  E(\lambda,B) = \sqrt{1+(\omega_c/2\omega_0)^2}E(\lambda')+M\omega_c/2\omega_0
  + g^{*}\mu_B B S_z/\hbar\omega_0
\end{equation}
and
\begin{equation}
  \lambda' = \lambda[1+(\omega_c/2\omega_0)^2]^{-1/4}
\end{equation}
where the symbol $\omega_c=eB/mc$ stands for cyclotron frequency. Note that there
is a misprint in (5.4) of \cite{matulis94}. The last term added in 
(\ref{energy_mag})
corresponds to the Zeeman energy where $g^{*}$ is the effective $g$ factor and 
$\mu_B$
is the Bohr magneton.

\section{Results and discussion}

As an illustration of the introduced renormalization technique we present
in figures \ref{fig3}--\ref{fig6} the excitation energy versus $\lambda$ plot
for the relative electron motion for the quantum dot with 2--5 electrons.
The inclusion of the center of mass motion energy is a trivial one. One should
just shift up the presented spectra by the some integer value, as the center of
mass motion is not affected by the electron interaction.

The main problem is to decide how the terms in both regions ($\lambda \to 0$ and
$\lambda \to \infty$) should be interconnected. Taking into account the fact
known in quantum mechanics that two terms with the same good quantum numbers
never cross each other we used the simplest term connection scheme.
Starting from the ground state we connected successively term in the region
$\lambda \to 0$ to the lowest possible unconnected term with the same $M$ and $S$
quantum numbers in the asymptotic region.

In figures \ref{fig3} and \ref{fig4} the energy plot for the system of two and
three electrons are presented. The terms with the zero energy exceeding the ground
state energy by 3 and 2, correspondingly, are depicted. In the case of four and 
five
electrons the energy plots are given in figures \ref{fig5} and \ref{fig6}.
The degeneracy of the states grows rapidly with the number of electrons and thus
here we present only the two lowest terms tending to each of asymptotic vibration
states. The asymptotic terms in the above figures are labeled by the numbers in
parenthesis along right axis which indicate the $k$ values of the phonons present
in the state. The phonon state with polarization corresponding to the larger
frequency is indicated by prime.

It can clearly be seen how the term structure transforms from that of the weakly
interacting quantum system, where the energy levels are governed by filling of
noninteracting single electron states to the semiclassical strongly correlated
system, where the energy levels correspond to the electronic ring vibration 
frequencies.
Such switching of the dot behaviour can be explained by simple dimensional 
considerations.
Namely, the three terms
in Hamiltonian (\ref{basic_ham}) can be estimated as $\hbar^2 N/mR^2$,
$m\omega_0^2R^2N$ and $e^2N(N-1)/\epsilon R$. Equating them we define the critical
$\lambda$ value $\lambda^{*} \sim 1/(N-1)$ where the above mentioned switching
occurs. This is in agreement with our energy dependencies.
As the dimensionless coupling constant $\lambda$ is proportional to
$a_0 \sim 1/\sqrt{\omega_0}$ the above estimation leads to
some condition for confinement potential frequency showing
whether the dot behaves like quantum system or the classical
one. When the confinement potential is mainly caused by some positive neutralizing
charge located on the rather distant gate the increase in confinement potential
frequency $\omega_0$  does not exceed $N$ \cite{haw93}. In that case increasing
of the number of electrons in a dot leads to more classical system.

As an illustration of the presented renormalized series technique
in the case with magnetic field (according to (\ref{energy_mag})) we calculated 
the
ground state energies vs. magnetic field for 2 -- 5 electrons which are
displayed in figures \ref{fig7} (a)--(d). The curve (a) for two electron system
is in good agreement with that presented in \cite{haw93} (we choose the parameters
as in \cite{haw93} and also include the fictitious gate charge but do not subtract
the zero point motion of the center of mass).
It is seen from figures \ref{fig7} (b)--(d) that increasing the number of 
electrons
the structure of the dependencies becomes less resolved and the peaks lose their
sharpness. Thus we miss a tiny additional peak in the position marked by an
arrow in figure \ref{fig7} (b) which is present in data of exact diagonalization.
Comparing our results with those obtained in \cite{maksym96}, say, for $N=3$,
$B = 20$ T, $S=3/2$, $M=3$ we obtained ground energy value $77.5$ meV
and find a good coincidence which proves that the approximate separation of 
rotation
and vibration modes works fairly well for small angular momenta.

Concluding we would like to mention that the parabolic quantum dot due to
the softness of the confinement potential is rather favourable object
for constructing the interpolating expressions in the contrary to the
hard wall dot \cite{matulis97} where the accuracy is much worse.

\ack

One of us (E. A.) is thankful to Prof. Per Hedeg\aa rd (\O rsted Laboratory,
University of Copenhagen) for useful comments.

\appendix

\section{Coulomb matrix element}\label{aps1}

In \cite{girvin83} the algebraic expression for calculation of the two-particle
matrix element of Coulomb interaction of electrons in the lowest Landau level
(corresponding to $n_i=0$ in our case) was used (see the comments in
\cite{stone92}).
In more general ($n_i \ne 0$) case the analogous Girvin and Jach expression can be
derived employing the same center of mass and relative motion complex coordinate
technique after expanding the associated Laguerre polynomials in sums and
representing the terms in form
\[
(r_1\e^{-i\varphi_1})^{\gamma_1}(r_2\e^{-i\varphi_2})^{\gamma_2}
(r_2\e^{i\varphi_2})^{\gamma_3}(r_1\e^{i\varphi_1})^{\gamma_4}
\frac{\e^{-(r_1^2+r_2^2)}}{|{\bi r}_1-{\bi r}_2|},
\]
here the symbols $\gamma_i$ stand for some numbers
\begin{eqnarray}\label{burbulas}
\gamma_1=j_1+j_4+(|m_1|+m_1)/2+(|m_4|-m_4)/2,\nonumber \\
\gamma_4=j_1+j_4+(|m_1|-m_1)/2+(|m_4|+m_4)/2.
\end{eqnarray}
Symbols $j_i$ are integer summation indices running form $0$ to $n_i$.
$\gamma_2$ and $\gamma_3$ can be obtained from \eref{burbulas}
replacing the indices $1\to 2$ and $4\to 3$.
The final expression reads
\begin{eqnarray}
\fl \langle 12|V|34\rangle =
\delta_{s_1,s_4}\delta_{s_2,s_3}\delta_{m_1+m_2,m_3+m_4}
  \left[\prod_{i=1}^4\frac{n_i!}{(|m_i|+n_i)!}\right]^{1/2}
  \sum_{(4)j=0}^{n}\frac{(-1)^{j_1+j_2+j_3+j_4}}{j_1!j_2!j_3!j_4!} \nonumber \\
  \times
  \left[\prod_{i=1}^4\left(\begin{array}{c} n_i+|m_i| \\
    n_i-j_i\end{array}\right)\right]
  \frac{1}{2^{(G+1)/2}}\sum_{(4)l=0}^{\gamma}
  (-1)^{\gamma_2+\gamma_3-l_2-l_3}\nonumber\\
  \times\delta_{l_1+l_2,l_3+l_4}\left[\prod_{i=1}^4\left(\begin{array}{c} \gamma_i
\\
    l_i\end{array}\right)\right]
  \Gamma\left(1+\frac{\Lambda}{2}\right)\Gamma\left(\frac{G-\Lambda+1}{2}
  \right)
  \label{stai}
\end{eqnarray}
where
\[
\fl
\sum_{(4)j=0}^{n}\equiv\sum_{j_1=0}^{n_1}\sum_{j_2=0}^{n_2}
\sum_{j_3=0}^{n_3}\sum_{j_4=0}^{n_4},\quad
G=\sum_i \gamma_i,\quad \Lambda=\sum_i l_i, \quad
\left( \begin{array}{c} n \\ m \end{array} \right) = \frac{n!}{m!(n-m)!}.
\]

Although the above expression involves terms with alternating signs and leads to
numerical difficulties in the case of large quantum number values \cite{stone92},
we found it rather convenient and accurate in the case of quantum dots with small
number of electrons.

\section{Expansion coefficients\label{coeff}}

Here the perturbation series (\ref{perturb}) coefficients $E_0$ and $E_1$,
and the inflection point angle $\phi$ are presented for all terms under
consideration. Those values together with the asymptotic expansion (\ref{eksp})
given in \tref{asfreq} and expressions (\ref{mapline}--\ref{rncoef}) are
sufficient for constructing the renormalized perturbation series.

\begin{table}
\caption{Coefficients for quantum dot with $N=2$ and $N=3$ electrons.}\label{du}
\begin{indented}\item[]
\begin{tabular}{@{}crcccccc}
\br
$N=2$ &&&& $N=3$ &&& \\
\mr
Term & $E_0$ & $E_1$ & $\phi$ & Term & $E_0$ & $E_1$ & $\phi$ \\
\mr
$^1S$   & 2 & 1.2533 & 0.3571 & $^2P$   & 4 & 2.8200 & 0.5871 \\
$^3P$   & 3 & 0.6267 & 1.2211 & $^4S$   & 5 & 1.8800 & 1.0722 \\
$^1D$   & 4 & 0.4700 & 1.3993 & $^2D$   & 5 & 2.2325 & 0.9358 \\
$^1S_0$ & 4 & 0.9400 & 0.7204 & $^2S_1$ & 5 & 2.5850 & 0.5836 \\
$^3F$   & 5 & 0.3917 & 1.4651 & $^4F$   & 6 & 1.7037 & 1.1984 \\
$^3P_0$ & 5 & 0.5483 & 1.3065 & $^4P_1$ & 6 & 1.8212 & 1.0783 \\
                           &&&& $^2P_1$ & 6 & 1.8413 & 1.0698 \\
                           &&&& $^2F_1$ & 6 & 2.2618 & 0.8991 \\
                           &&&& $^2P_0$ & 6 & 2.5942 & 0.7067 \\
\br
\end{tabular}
\end{indented}
\label{tap1}
\end{table}

\begin{table}
\caption{Coefficients for quantum dot with $N=4$ and $N=5$
electrons.}\label{penki}
\begin{indented}\item[]
\begin{tabular}{@{}crcccccc}
\br
$N=4$ &&&& $N=5$ &&& \\
\mr
Term & $E_0$ & $E_1$ & $\phi$ & Term & $E_0$ & $E_1$ & $\phi$ \\
\mr
$^3S$       & 6 & 5.0133 & 0.6105 & $^2P$   &  8 & 8.3032 & 0.4913 \\
$^1D$       & 6 & 5.2483 & 0.5800 & $^4D$   &  9 & 7.4416 & 0.6649 \\
$^1S$       & 6 & 5.4832 & 0.5527 & $^2S$   &  9 & 7.4812 & 0.6607 \\
$^3P$       & 7 & 4.3608 & 0.8333 & $^4S_2$ &  9 & 7.6765 & 0.6409 \\
$^3F$       & 7 & 4.6412 & 0.7828 & $^2S_2$ &  9 & 7.7647 & 0.6325 \\
$^1P_1$     & 7 & 4.6999 & 0.6087 & $^2G$   &  9 & 7.8234 & 0.6270 \\
$^3P_2$     & 7 & 4.8628 & 0.6482 & $^2D$   &  9 & 7.8284 & 0.6265 \\
$^5D$       & 8 & 3.5641 & 1.0635 & $^2D_2$ &  9 & 8.1710 & 0.5969 \\
$^5S_2$     & 8 & 3.6816 & 0.9920 & $^2S_2$ &  9 & 8.3125 & 0.5263 \\
$^3S_1$     & 8 & 3.9491 & 0.8956 & $^4P$   & 10 & 6.7149 & 0.8303 \\
$^1S_2$     & 8 & 3.9822 & 0.9218 & $^4F$   & 10 & 6.8596 & 0.8110 \\
$^3D_2$     & 8 & 4.0215 & 0.9130 & $^4P_2$ & 10 & 6.9609 & 0.7980 \\
$^1G$       & 8 & 4.0919 & 0.9543 & $^2F$   & 10 & 6.9615 & 0.7980 \\
$^3G$       & 8 & 4.1320 & 0.9466 & $^2P_2$ & 10 & 7.1355 & 0.7768 \\
$^1D_2$     & 8 & 4.3311 & 0.8487 & $^4F_2$ & 10 & 7.2010 & 0.7224 \\
$^3D_1$     & 8 & 4.3783 & 0.8004 & $^2P_2$ & 10 & 7.2489 & 0.7168 \\
$^1S_{2'}$  & 8 & 4.3866 & 0.7810 & $^2H$   & 10 & 7.2543 & 0.7631 \\
$^3S_1$     & 8 & 4.5628 & 0.7656 & $^2F_2$ & 10 & 7.3103 & 0.7568 \\
$^3D_1$     & 8 & 4.5642 & 0.7654 & $^4P_2$ & 10 & 7.3563 & 0.7046 \\
$^1G_2$     & 8 & 4.6678 & 0.7885 & $^2P_1$ & 10 & 7.4043 & 0.6993 \\
$^1D_{2'}$  & 8 & 4.7087 & 0.7220 & $^2P_1$ & 10 & 7.5375 & 0.6783 \\
$^3S_0$     & 8 & 4.7653 & 0.6836 & $^2F_2$ & 10 & 7.6303 & 0.6757 \\
$^1S_0$     & 8 & 4.8182 & 0.6751 & $^2F_1$ & 10 & 7.6670 & 0.6720 \\
$^1D_0$     & 8 & 5.0062 & 0.6472 & $^2P_{2,2}$ & 10 & 7.9283 & 0.6407 \\
$^1S_{2,2}$ & 8 & 5.2186 & 0.6140 & $^2P_{2,2}$ & 10 & 8.0142 & 0.6331 \\
\br
\end{tabular}
\end{indented}
\label{tap2}
\end{table}

\Figures

\begin{figure}
\caption{Mapping in the $\xi\lambda$-parameter plane.}
\label{fig1}
\end{figure}
\begin{figure}
\caption{The first excited triplet state for two electrons:
1, Energy versus $\phi$ plot;
2, the exact result;
3, energy value at the inflection point.}
\label{fig2}
\end{figure}
\begin{figure}
\caption{Spectrum for two electrons.}
\label{fig3}
\end{figure}
\begin{figure}
\caption{Spectrum for three electrons.}
\label{fig4}
\end{figure}
\nopagebreak
\begin{figure}
\caption{Spectrum for four electrons.}
\label{fig5}
\end{figure}
\begin{figure}
\caption{Spectrum for five electrons.}
\label{fig6}
\end{figure}
\begin{figure}
\caption{Ground state energy of quantum dot in magnetic field:
(a) - 2 electrons, (b) - 3 electrons, (c) - 4 electrons, (d) - 5 electrons.
Single electron ground state energies $E(1)$ are subtracted.}
\label{fig7}
\end{figure}
%

\end{document}